\documentclass[prb,twocolumn,notitlepage,superscriptaddress]{revtex4-2}
\usepackage{amsmath,amsfonts,amssymb,amsthm,epsfig,array}
\usepackage{dsfont}
\usepackage{slashed}
\usepackage{graphics}
\usepackage{float}
\usepackage{verbatim}
\usepackage{color}
\usepackage[dvipsnames]{xcolor}
\usepackage[hidelinks,colorlinks,linkcolor=blue,
citecolor=blue,urlcolor=blue]{hyperref}
\usepackage[titletoc,title]{appendix}
\usepackage{multirow}
\usepackage{bibentry}

\newcommand{\bsl}[1]{\boldsymbol{#1}}

\newcommand{\ii}{\mathrm{i}}

\newcommand{\bra}[1]{\langle #1|}
\newcommand{\ket}[1]{|#1 \rangle}

\newcommand{\dsZ}{\mathbb{Z}}

\newcommand{\U}[1]{\mathrm{U}(#1)}

\newcommand{\eqnref}[1]{Eq.\,\eqref{#1}}
\newcommand{\figref}[1]{Fig.\,\ref{#1}}

\newcommand{\secref}[1]{Sec.\,\ref{#1}}

\newcommand{\refcite}[1]{Ref.\,[\onlinecite{#1}]}

\newcommand{\eq}[1]{\begin{equation} #1 \end{equation}}

\newcommand{\eqa}[1]{\begin{align}\begin{split} #1 \end{split}\end{align}}
\let\oldAA\AA
\renewcommand{\AA}{\text{\normalfont\oldAA}}
\newcommand{\sgn}[1]{\text{sgn}(#1)}

\newcommand{\ie}{{\emph{i.e.}}}
\newcommand{\eg}{{\emph{e.g.}}}
\newcommand{\TR}{\mathcal{T}}
\newcommand{\cc}{\mathcal{K}}
\usepackage[mathscr]{euscript}

\newcommand{\si}{\sigma}

\newcommand{\mbt}{{MnBi$_2$Te$_4$}}

\usepackage[normalem]{ulem}

\begin{document}
\title{Pseudo-gauge Fields in Dirac and Weyl Materials}
\author{Jiabin Yu}
\affiliation{Condensed Matter Theory Center, Department of Physics, University of Maryland, College Park, Maryland 20742, USA}
\author{Chao-Xing Liu}
\email{cxl56@psu.edu}
\affiliation{Department of Physics, the Pennsylvania State University, University Park, PA 16802}
\begin{abstract}
Electrons in low-temperature solids are governed by the non-relativistic Schr$\ddot{o}$dinger equation, since the electron velocities are much slower than the speed of light. 
Remarkably, the low-energy quasi-particles given by electrons in various materials can behave as relativistic Dirac/Weyl fermions that obey the relativistic Dirac/Weyl equation.
These materials are called ``Dirac/Weyl materials", which provide a tunable platform to test relativistic quantum phenomena in table-top experiments.
More interestingly, different types of physical fields in these Weyl/Dirac materials, such as magnetic fluctuations, lattice vibration, strain, and material inhomogeneity, can couple to the ``relativistic" quasi-particles in a similar way as the $U(1)$ gauge coupling. 
As these fields do not have gauge-invariant dynamics in general, we refer to them as “pseudo-gauge fields”. 
In this chapter, we overview the concept and the physical consequences of pseudo-gauge fields in Weyl/Dirac materials.
In particular, we will demonstrate that pseudo-gauge fields can provide a unified understanding of a variety of physical phenomena, including chiral zero modes inside a magnetic vortex core of magnetic Weyl semimetals, a giant current response at magnetic resonance in magnetic topological insulators, and piezo-electromagnetic response in time-reversal invariant systems.
These phenomena are deeply related to various concepts in high-energy physics, such as chiral anomaly and axion electrodynamics. 
\end{abstract}
\maketitle

\section{Introduction}
The concepts of Dirac fermions and gauge fields lie at the heart of physics~\cite{Srednicki2007QFT}.
In the Standard Model of particle physics, all fermions---except neutrinos---are confirmed to be Dirac fermions, and the interactions among Dirac fermions are mediated by gauge fields.
One prominent part of the Standard Model is the theory of quantum electrodynamics, which is one of the most precisely tested theories up to now. 
Specifically, the theory treats electrons as Dirac fermions that obey the massive relativistic Dirac equation, and describes the electromagnetic interaction among electrons by the $\U{1}$ gauge field.
Conceptually, limiting the mass in the Dirac equation to zero allows us to block-diagonalize the Dirac equation into two decoupled massless Weyl equations, indicating that a massless Dirac fermion consists of two massless Weyl fermions.
Interestingly, Weyl fermions are only allowed to appear in pairs, since the appearance of a single Weyl fermion breaks the charge conservation quantum mechanically---having the so-called chiral anomaly~\cite{Nielsen1981ChialFermionDoubling,Srednicki2007QFT}.

In condensed matter physics, we are often interested in behaviors of electrons.
However, we do not need to use relativistic Dirac equations to describe electrons in condensed matter systems, since their velocities are much smaller than the speed of light; instead, we can use the non-relativistic Schr$\ddot{o}$dinger equation.
Moreover, the number of electrons in real materials is very large $(\sim 10^{23})$, and they are subject to the background potential (\eg, lattice potential) and interact with each other, making directly solving the Schr$\ddot{o}$dinger equation extremely difficult.
Thankfully, owing to the Pauli principle, only the electrons near the Fermi energy dominate the behaviors of the systems in equilibrium at low temperatures.
It turns out that in many cases, those low-energy electrons behave as  quasi-particles that obey different types of equations of motions.

Intriguingly, in certain condensed matter systems, the low-energy quasi-particles obey noninteracting relativistic Dirac or Weyl equations, and thus these material systems are respectively referred to as Dirac or Weyl materials~\cite{Yang2016DiracWeyl}. 
The most prominent examples include the spatially-two-dimensional (2D) massless Dirac fermions in graphene~\cite{Neto2009GrapheneRMP,Geim2010RiseOfGraphene} and at the surface of topological insulators (TIs)~\cite{Hasan2010TI,Qi2010TITSC}, and spatially-three-dimensional (3D) massless Dirac/Weyl fermions in Dirac/Weyl semimetals~\cite{Wan2011WSM,Fang2012DSM,Hosur2013WSM,Yan2017WSM,Armitage2018RMPWeylDirac}.
As Dirac/Weyl quasi-particles are provided by electrons, they carry nonzero electric charge and couples to the electromagnetic $\U{1}$ gauge field.
Consequently, these materials serve as fertile and controllable platforms for exploring gauge-field-induced exotic relativistic phenomena in table-top experiments. 
For example, various signatures of the chiral anomaly of Weyl quasi-particles have been predicted or even experimentally detected in Dirac/Weyl materials~\cite{Hosur2013WSM,Yan2017WSM,Armitage2018RMPWeylDirac}, such as negative magneto-resistance~\cite{Nielsen1983ABJWeyl, Huang2015WSMChiralAnomaly,Xiong2015DSMChiralAnomaly,Zhang2016WSMchiralanomaly,Ma2017chiralityWSM} and the large magneto-optical Kerr effect~\cite{Hosur2015ChiralAnomalyWeyl}.

Besides the $\U{1}$ gauge field, Dirac/Weyl quasi-particles can couple to the so-called ``pseudo-gauge field"~\cite{Ilan2020NatRevPseudoGauge}.
A pseudo-gauge field is a field that (i) couples to one Dirac/Weyl quasi-particle in the same way as the $\U{1}$ gauge field but (ii) does not have gauge-invariant dynamics.
Owing to the linear-in-momentum form of the Hamiltonian of Dirac/Weyl quasi-particles, a variety of physical fields---including magnetization, strain, and phonons---can be treated as pseudo-gauge fields. 
Interestingly, unlike the $\U{1}$ gauge field that typically has the same form for all Dirac/Weyl quasi-particles, pseudo-gauge fields can have different forms for (or in other words couple differently to) different Dirac/Weyl quasi-particles. 
As a result, in Dirac/Weyl materials with multiple Dirac/Weyl quasi-particles, a pseudo-gauge field may induce unusual physical responses and phenomena that cannot be induced by the usual electromagnetic field.  

In this chapter, we will review the occurrence of pseudo-gauge fields in Dirac/Weyl materials and the resultant nontrivial responses.
We will first discuss the general mechanism for the occurrence of pseudo-gauge fields in Dirac/Weyl materials in \secref{sec:gen_mech}.
Then, we will focus on the pseudo-gauge fields that are induced by the magnetic vortices, the magnetic fluctuations, and the strain (or similarly phonons).
In \secref{sec:mag_pseudogauge}, we will discuss chiral zero Landau levels located at the magnetic vortex cores in magnetic Weyl semimetals, and the giant current response induced by magnetic resonance in axion insulators.
In \secref{sec:strain_pseudogauge}, we will discuss the strain-induced current response (piezoelectric effect) that is discontinuous across a 2D time-reversal (TR) invariant topological quantum phase transition, and the strain-induced bulk orbital magnetization in a 3D TR-invariant Weyl semimetal gapped by a charge-density wave (CDW).
In this book chapter, we will mainly review our works on the pseudo-gauge fields in Dirac/Weyl materials, while many other works have been comprehensively reviewed in \refcite{Ilan2020NatRevPseudoGauge}.
We emphasize that the field of pseudo-gauge fields is developing rapidly, and an outlook of this field will be presented in \secref{sec:conclusion}. 

\section{Dirac Quasi-particles and Emergent Pseudo-Gauge Fields}
\label{sec:gen_mech}

In this section, we first review the general framework of pseudo-gauge field in effectively noninteracting Dirac/Weyl materials. 
Here we mainly focus on the Dirac fermions, while the Weyl fermions can be obtained by setting the mass $m$ in the Dirac equations to be zero since all Weyl fermions must appear in pairs in Weyl semimetals. 
In $d$ spatial dimensions and one time dimension, the Lagrangian for a Dirac fermion coupled to the $\U{1}$ gauge field reads~\cite{Srednicki2007QFT}
\eq{
\label{eq:Dirac_Fermion}
\mathcal{L}= \overline{\psi} [\ii (\slashed{\partial}-\ii q \slashed{A})-m]\psi\ ,
}
where $\psi$ is the Dirac field, $A$ is the $\U{1}$ gauge field, and $m$ and $q$ are respectively the mass and charge of the Dirac fermion.
Throughout this chapter, we adopt the unit system in which $\hbar=c=1$.
The slash is defined as $\slashed{a}=\gamma^{\mu}a_{\mu}$ with Einstein summation notation always implied for Greek letter indices, $\mu=0,1,2,...d$ with $0$ for the time dimension, and $\gamma^{\mu}$ are gamma matrices that satisfy $\{\gamma^{\mu},\gamma^{\nu}\}=-2g^{\mu\nu}$ with $g=diag(-,+,...,+)$ the metric.
In particular, $\psi$ has two components in $2+1D$ and four components in $3+1$D, and $\overline{\psi}=\psi^\dagger \gamma^0$.
Note that throughout this chapter, we use $d$D for $d$ spatial dimensions, and use $d+1$D for $d$ spatial dimensions plus one time dimension.
With this convention, $d$D and $d+1$D have the same physical meaning, and thus can be used interchangeably.

In a Dirac material, there may be more than one Dirac quasi-particles, and thereby the Lagrangian for them (in the presence of $\U{1}$ gauge field) reads 
\eq{
\label{eq:Dirac_QP_A}
\mathcal{L}= \sum_{i}\overline{\psi}_a \left[\ii (\slashed{\partial}+\ii e \slashed{A})-m_a e^{-\ii \theta_a \gamma^5}\right]\psi_{a}\ ,
}
where $a$ labels different Dirac quasi-particles, and we have substitute $q=-e$ for the Dirac quasi-particles with $-e$ the electron charge.
Here we include $\gamma^5$ which is defined as 
\eqa{
& \gamma^5 = 0 \text{ for 2+1D}\\
& \gamma^5 = \ii \gamma^0 \gamma^1 \gamma^2 \gamma^3 \text{ for 3+1D}\ ,
}
where we choose $\gamma^5$ to be zero for $2+1$D because all three Pauli matrices are included in $\gamma^\mu$ for $2+1$D.
The reason for including $\gamma^5$ is that the phase of the mass $\theta_a$ is generally allowed for Dirac quasi-particles in $3+1$D, owing to the lower symmetries for a single Dirac quasi-particle compared to a Dirac fermion.
Owing to the nonzero $\gamma^5$ in $3+1D$, we choose $m_a=|m_a|$ in $3+1$D.
\eqnref{eq:Dirac_QP_A} suggests that the $\U{1}$ gauge field has the universal coupling $e$ to all Dirac quasi-particles.

Let us now include a generic physical field $\xi$ that has the following leading-order coupling to the Dirac quasi-particles 
\eq{
\label{eq:Dirac_QP_xi}
\mathcal{L}_{\psi\xi}= -\sum_{a,l}\overline{\psi}_a \Lambda_a^l\psi_a \xi_l\ ,
}
where $l$ is the component index of $\xi$ and $\Lambda_a^l$ are matrices.
The $\Lambda_a^l$ matrices can always be separated into the gamma-matrix part and other part as
\eq{
\Lambda_a^l=\lambda_{a,\mu}^l\gamma^\mu+ \kappa_{a,\mu}^l\gamma^\mu \gamma^5 + ...
}
If the ``..." part is negligible, we can define
\eqa{
& A^{pse}_{a,\mu}=\frac{1}{e}\sum_{l}\lambda_{a,\mu}^l\xi_l \\
& A_{5, a,\mu}=\frac{1}{e}\sum_{l}\kappa_{a,\mu}^l\xi_l
}
and combining the \eqnref{eq:Dirac_QP_xi} with \eqnref{eq:Dirac_QP_A} gives
\eq{
\label{eq:Dirac_QP_A_Apse}
\mathcal{L}= \sum_{a}\overline{\psi}_a [\ii (\slashed{\partial}+\ii e \slashed{A}+\ii e \slashed{A}^{pse}_{a}+\ii e \slashed{A}_{5,a} \gamma^5)-m_a e^{-\ii \theta_a \gamma^5}]\psi_a\ .
}
\eqnref{eq:Dirac_QP_A_Apse} suggests that $A^{pse}_a$ provided by the physical field $\xi$ is a pseudo-gauge field, since $A^{pse}_a$ couples to $a$th Dirac quasi-particle in the same way as the $\U{1}$ gauge field.
Moreover, \eqnref{eq:Dirac_QP_A_Apse} also suggests that the physical field $\xi$ can also provide the chiral gauge field  $A_{5,a}$ for Dirac quasi-particles, which has not been experimentally found for the Dirac fermions in high-energy physics.
In fact, when the Dirac quasi-particles have zero mass, the chiral gauge field $A_{5,a}$ becomes the pseudo-gauge field for each individual Weyl quasi-particle, and therefore in general, we can also refer to $A_{5,a}$ as a pseudo-gauge field.

The reason for having pseudo-gauge fields in \eqnref{eq:Dirac_QP_A_Apse} at the leading order is that Dirac (and Weyl) quasi-particles have linear-in-momentum Lagrangian to the leading order and thereby linearly couple to the $\U{1}$ gauge field.
Thus, including pseudo-gauge fields in Dirac materials is quite natural in general. 
In contrast, if the fermion Lagrangian to the leading order was quadratic in momentum, the leading order perturbation could not play the role of gauge fields.
In this case, the higher-order coupling, as well as the fine tuning of the coupling parameters, would be needed to provide pseudo-gauge fields, which is not natural in real materials. 

The field strength of the $\U{1}$ gauge field is $F_{\mu\nu}=\partial_{\mu} A_{\nu}-\partial_{\nu} A_{\mu}$.
In analogy, we can define the field strength for pseudo-gauge field as $F^{pse}_{\mu\nu}=\partial_{\mu} A_{\nu}^{pse}-\partial_{\nu} A_{\mu}^{pse}$, from which we can define the pseudo-electric and pseudo-magnetic fields as 
\eqa{
& \bsl{E}^{pse}=(F^{pse}_{10}, F^{pse}_{20}, F^{pse}_{30}) \\
& \bsl{B}^{pse}=(F^{pse}_{23}, F^{pse}_{31}, F^{pse}_{12})\ 
}
for $3+1D$.
The fields for $2+1$D can be obtained by excluding all $F^{pse}$ elements with $\mu=3$ or $\nu=3$.
In general, when discussing the response induced by pseudo-gauge field, we actually mean the response directly induced by pseudo-electric and pseudo-magnetic fields.
For example, in 2+1D (implying $\gamma^5=0$), we can directly integrate out the fermions in \eqnref{eq:Dirac_QP_A_Apse} and get the following effective Chern-Simons Lagrangian~\cite{TKNN,Zhang1992ReviewChernSimons}
\eq{
\mathcal{L}^{eff}_{2+1D}=\sum_{a} \frac{\sigma_a}{2}\epsilon^{\mu\nu\rho} \widetilde{A}_{a,\mu}\partial_\nu \widetilde{A}_{a,\rho}+...
}
where $\sigma_a=-\frac{\sgn{m_a}e^2}{4\pi}$ is the Hall conductance of each Dirac quasi-particle, $\widetilde{A}_{a,\mu}= A_\mu+ A^{pse}_{a,\mu}$, and ``$...$" contains all other terms.
From the above expression, we can derive the response current as $J_\mu=J_\mu^{phys}+J^{pse}_\mu+...$, where 
\eqa{
& \rho^{phys}=-J_0^{phys}=\sum_{a} \sigma_a B_z\\
& J_i^{phys}=\sum_{a,j} \sigma_a \epsilon^{i j} E_j\\
& \rho^{pse}=-J^{pse}_0=\sum_{a} \sigma_a B_{a,z}^{pse}\\
& J_i^{pse}=\sum_{a,j} \sigma_a \epsilon^{i j} E_{a,j}^{pse}\ ,
}
$\rho=J^0$, and $i,j\in \{x,y\}$.
Here the superscripts $phys$ and $pse$ just label the current contributions that come from the physical gauge fields and the pseudo-gauge fields, respectively; both $J_\mu^{phys}$ and $J^{pse}_\mu$ are parts of the physical response current $J^\mu$.
We can see the pseudo-electric field can induce the Hall current, and the pseudo-magnetic field can change the charge density.

In 3+1D (implying $m_a=|m_a|$) with $A_{5,a}=0$ in \eqnref{eq:Dirac_QP_A_Apse}, the pseudo-electric field can still induce the 2+1D effect as discussed above.
In addition, as we allow the mass of the Dirac fermion to be complex, we will have a new response---effective axion electrodynamics.
To be specific, let us integrate out the fermions, leading to the following effective Lagrangian~\cite{Qi2008TFT}
\eq{
\label{eq:S_eff_a_gen}
\mathcal{L}_{eff}= \sum_{a} \frac{e^2}{16\pi^2} \frac{\theta_a}{2} \varepsilon^{\mu\nu\rho\delta}  \widetilde{F}_{a,\mu\nu}\widetilde{F}_{a,\rho\delta}+ ...\ ,
}
where $\widetilde{F}_{a,\mu\nu}=\partial_\mu \widetilde{A}_{a,\nu}-\partial_\nu \widetilde{A}_{a,\mu}$.
The leading term in \eqnref{eq:S_eff_a_gen} has the same form as the axion electrodynamics~\cite{Qi2008TFT}, and thus $\theta_a$ is called the effective axion field.
The response derived from the leading term in $\mathcal{L}_{eff,a}$ reads
\eqa{
\label{eq:TME_bulk}
& \bsl{P}= -\frac{e^2}{2\pi}\sum_a\frac{\theta_a}{2\pi}(\bsl{B}+\bsl{B}^{pse}_a)\\
& \bsl{M}= -\frac{e^2}{2\pi}\sum_a\frac{\theta_a}{2\pi}(\bsl{E}+\bsl{E}^{pse}_a)\ ,
}
where $\bsl{P}$ and $\bsl{M}$ are the polarization and magnetization of the sample, respectively.
We can see the pseudo-gauge field in this case can contribute to both $\bsl{P}$ and $\bsl{M}$, through the effective axion field.
When the pseudo-gauge field is omitted and $\sum_{a}\theta_a$ is nonzero and quantized, \eqnref{eq:TME_bulk} represents the topological magnetoelectric effect~\cite{Qi2008TFT}.

When both $A_{a,5}$ and $m_a$ are zero in \eqnref{eq:Dirac_QP_A_Apse}, the action describes the effective theory of Weyl semimetals and has chiral anomaly~\cite{Bertlmann2000Anomalies,Srednicki2007QFT}.
Specifically, the chiral current $ J^\mu_{5,a}= \langle\bar{\psi}_a \gamma^\mu\gamma^5 \psi_a\rangle$ of each massless Dirac quasi-particle (or equivalently the difference between currents of Weyl quasi-particles with opposite chiralities) is not conserved:
\eq{
\partial_\mu J^\mu_{5,a}= -\frac{e^2}{16\pi^2} \varepsilon^{\mu\nu\rho\delta}  \widetilde{F}_{a,\mu\nu}\widetilde{F}_{a,\rho\delta}\ ,
}
where $\langle ... \rangle$ represents the average.
This equation suggests that the pseudo-gauge field can also contribute to the chiral anomaly of the system.

In practice, various fields---such as magnetization, strain and phonons---can couple to the Dirac quasi-particles in the form of \eqnref{eq:Dirac_QP_xi}, and thereby can produce pseudo-gauge fields.
In particular, the pseudo-gauge fields can have different forms for different Dirac fermions, allowing the pseudo-gauge fields to induce exotic responses that are forbidden for the $\U{1}$ gauge field.
In the following, we provide several examples of pseudo-gauge fields and discuss the exotic response induced by them.

\section{Magnetization-induced Pseudo-gauge field in Magnetic Topological Systems}
\label{sec:mag_pseudogauge}

3D TIs possess an insulating bulk gap and gapless surface states. The minimal model for the bulk states of TIs can be described by a massive 3D Dirac Hamiltonian, while the topological surface states behave as 2D massless Dirac quasi-particles.
Certain configurations of magnetic TIs and trivial insulators (such as a superlattice with the alternating magnetic TI layers and trivial insulator layers) can realize magnetic Weyl semimetal phases, even with the minimal two Weyl points~\cite{Burkov2011WSMTILayers}.
Thus, 3D TIs provide a desirable platform to explore the pseudo-gauge field effect. 
This section focuses on the effect induced by magnetization in topological systems, and particularly, we will show that magnetic vortex structure can create a pseudo-magnetic field while magnetic dynamics can induce a pseudo-electric effect. These pseudo-gauge field effects can bring us intriguing theoretical predictions, including chiral zero modes located at the magnetic vortex cores in magnetic Weyl semimetals, and a giant current response at magnetic resonance in magnetic TIs. 

\subsection{Chiral Zero Modes at Magnetic Vortex cores in Magnetic Weyl semimetals}
\label{sec:Vortex_MTI}

In this part, we will first discuss the physical phenomena induced by the magnetic texture, particularly magnetic vortex, in magnetic Weyl semimetals.
Here we follow \refcite{Liu2013WSMChiralGF} to consider a minimal TI model with a uniform magnetization to realize the Weyl semimetal phase. 

We start from the minimal four-band model for TIs, which is given by
$H_0=\mathcal{M}({\bf k})\tau_z+A_1 k_z\tau_y+A_2(k_y\sigma_x-k_x\sigma_y)\tau_x$, 
where $\mathcal{M}({\bf k})=M_0+M_1 k_z^2+M_2 (k_x^2+k_y^2)$, $\tau$ and $\sigma$ are two sets of Pauli matrices for orbitals (with opposite parities) and spins, respectively. $M_0, M_1, M_2, A_1$ and $A_2$ are material dependent parameters. With magnetic doping, the exchange coupling between the z-direction magnetization and electron spin includes two possible terms $H_1=U_z\sigma_z+V_z\sigma_z\tau_z$ since there are two orbitals in our model and the exchange coupling constants for different orbitals can be different. However, for simplicity, we only keep the $U_z$ term below ($V_z=0$). For the Hamiltonian $H_0+H_1$, when the condition $|U_z|>|M_0|$ is satisfied, the system will be in the Weyl semimetal phase with the Weyl points located at two momenta ${\bf k}_{W,\pm}=(0,0,\pm K_0)$ with $K_0=\frac{1}{A_1}\sqrt{U_z^2-M_0^2}$. The low-energy physics of this system is described by two Weyl quasi-particles with opposite chiralites, or equivalently one massless Dirac quasi-particle.

\begin{figure*}[t]
    \centering
    \includegraphics[width=1.6\columnwidth]{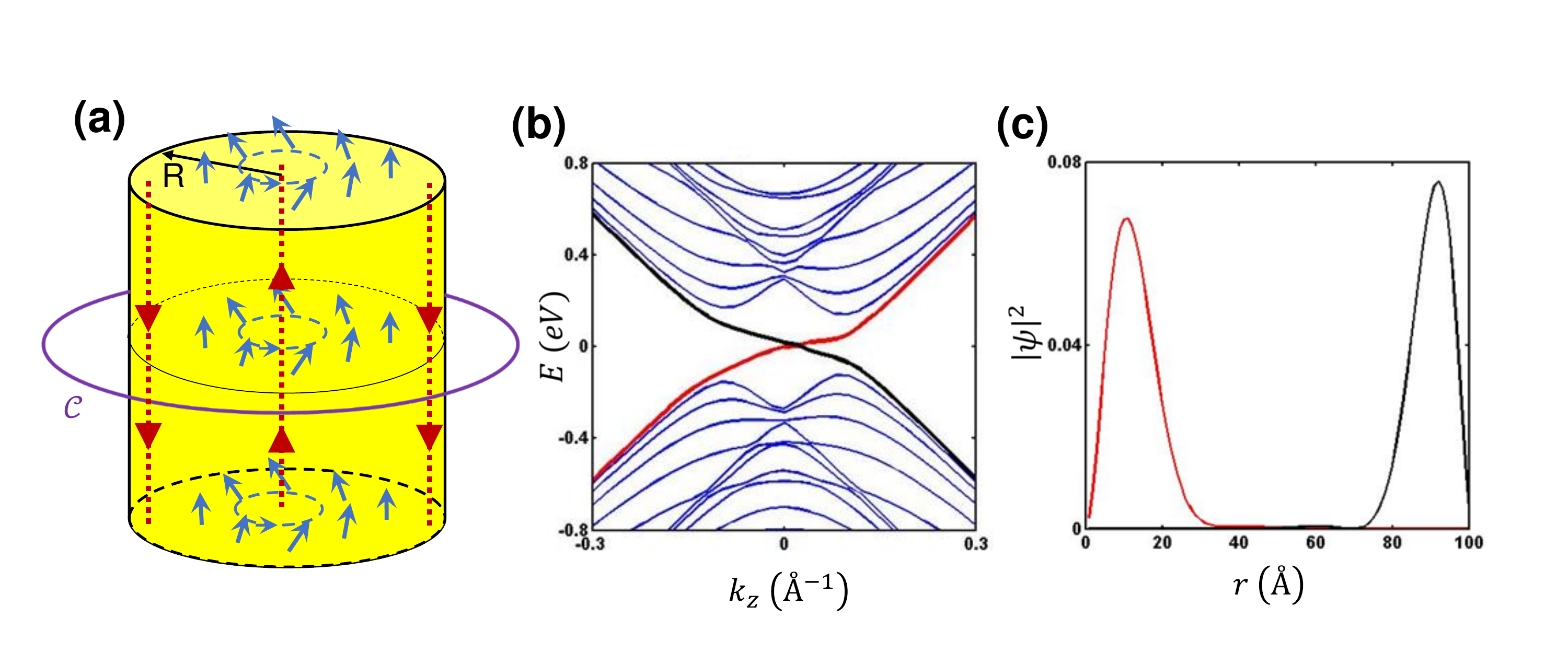}
    \caption{(Adapted from \refcite{Liu2013WSMChiralGF}.) (a) Schematics of a cylinder of a magnetic Weyl semimetal with magnetic vortex. Here the blue arrows depict magnetic texture, the red dashed lines with the red arrows show the chiral zero modes located at the vortex core and the surface, and $R$ is the radius of the cylinder. The loop $\mathscr{C}$ encloses the cylinder and the integral of the ${\bf a}$ field along the loop $\mathscr{C}$ is zero. (b) shows the energy spectrum for the cylinder of magnetic Weyl semimetal with magnetic vortex. (c) shows the wavefunctions of two chiral zero modes in (b) and the x-axis $r$ is the radius.  }
    \label{fig:mag_vort}
\end{figure*}

We next study the influence of magnetic texture on top of z-direction magnetization in the Weyl semimetal phase, as shown in \figref{fig:mag_vort}a and consider the Hamiltonian term $H_2=\sum_{i=x,y,z}(\mu_i\sigma_i+\nu_i\sigma_i\tau_z)$ to describe the coupling between electron spin and magnetic texture. Here both $\mu_i$ and $\nu_i$ ($i=x,y,z$) are the coupling parameters for the magnetization along the $i$-direction. We can project the whole system ($H_0+H_1+H_2$) into the low-energy subspace of two Weyl quasi-particles, and the effective Hamiltonian reads (see the detailed derivation in \refcite{Liu2013WSMChiralGF}) 
\eq{
H_{eff}=v_f\left(\begin{array}{cc} ({\bf k}+e{\bf A}+{\bf a})\cdot{\bf \sigma} & 0\\0 & -({\bf k}+e{\bf A}-{\bf a})\cdot{\bf \sigma}
\end{array}\right), 
}
under appropriate basis, where ${\bf a}$ is given by $a_z=\frac{1}{v_f}\mu_z$, $a_x=\frac{A_1K_0}{v_fU_z}\nu_x$ and $a_y=\frac{A_1 K_0}{v_f U_z}\nu_y$ up to the linear order in $\mu_i$ and $\nu_i$, from which one can see that ${\bf a}$ is directly determined by magnetic texture. The Pauli matrices $\sigma$ here are for spin while two blocks are for the expansion around ${\bf k}_{W,\pm}$. Here we have re-scaled the momentum along different directions to make the dispersion isotropic with one Fermi velocity $v_f$, and also include the coupling to the vector potential ${\bf A}$ for electromagnetic fields for comparison. Now one can clearly see that if we treat ${\bf a}$ as a field with smoothly spatial and temporal dependence, it plays a similar role as ${\bf A}$ for each Weyl quasi-particle, but as the sign of the gauge coupling is opposite for two Weyl quasi-particles, ${\bf a}$ is actually an chiral gauge field for the Dirac quasi-particle formed by the two Weyl quasi-particles. This model gives us a simple example to show how magnetization can play the role of pseudo-gauge field. 

This understanding also implies the existence of chiral modes in the magnetic vortex core. To see that, let us first assume the configuration ${\bf a}=b_0(-\frac{y}{2},\frac{x}{2},0)$ for the ${\bf a}$ field induced by certain type of magnetic texture, where $b_0$ is a parameter. This ${\bf a}$ field configuration corresponds to the symmetric gauge for a uniform pseudo-magnetic field ${\bf b}=\nabla\times {\bf a}=b_0 \hat{e}_z$ along the z-direction. This is just the Landau level problem for Dirac fermions and the Landau level spectrum is given by $E_{\pm,\alpha}(n)=v_f\sqrt{k_z^2+2b_0n}$ with $n=1,2,3,...$ and $\alpha=\pm$ for two Weyl quasi-particles with opposite chiralites, as well as the additional zero mode solutions $E_{\alpha}(n=0)=-v_f k_z$. We notice that both $\alpha=\pm$ zero modes have the same velocity along the z-direction, and thus chiral modes will appear in the magnetic vortex core. 

However, the above simple picture is neither complete nor realistic. For the mentioned ${\bf a}$ configuration, the amplitude of ${\bf a}$ will increase with the radius $r=\sqrt{x^2+y^2}$, while in any real system, the magnetization should not exceed certain value, and thus the amplitude of ${\bf a}$ field should approach a constant for a large $r$. Consequently, the pseudo-magnetic field ${\bf b}$ is not uniform, but mainly concentrates around the magnetic vortex core. Furthermore, for a finite sample, the magnetization vanishes outside the sample and that means the loop integral over the ${\bf a}$ field along the loop $\mathscr{C}$ shown in \figref{fig:mag_vort}a should be zero, or equivalently this loop should enclose total zero flux for the pseudo-magnetic field ${\bf b}$. To solve these problems and study more realistic systems, we consider a finite cylinder with a magnetic vortex around the cylinder center and numerically solved the model Hamiltonian $H_0+H_1+H_2$ in this finite system (see \refcite{Liu2013WSMChiralGF}). The energy spectrum in \figref{fig:mag_vort}b shows two counter-propagating zero modes and the wavefunctions of these two modes are shown in \figref{fig:mag_vort}c, from which one can see that one of these two chiral modes is located at the vortex core while the other appears at the outer surface of the cylinder, which is also schematically depicted in \figref{fig:mag_vort}a. When these two counter-propagating zero modes are spatially separated far enough, the mixing between these two modes will be negligible, similar to the case of chiral edge states of the quantum Hall effect. 

The presence of the chiral zero Landau levels implies the quantum anomaly, which means that the charge non-conservation due to the quantum correction. Since magnetization serves as the chiral gauge field in this system, there is the subtlety if the anomaly equations will involve the U(1) charge non-conservation or not, which is known as the consistent and covariant anomalies in the context of high energy physics. Physically, the non-conservation of U(1) charge is related to the fact that some high energy parts of the bands in Weyl semimetals also carry the Berry curvature and thus contribute to the Hall current response, but these contributions are not included in the low energy physics of Weyl quasi-particles. This causes the complexity in writing down the anomaly equations for Weyl semimetals. A comprehensive summary of this issue can be found in the review paper \refcite{Ilan2020NatRevPseudoGauge} and the references therein. 

Magnetic Weyl semimetal phase has been experimentally realized in Mn$_3$Sn~\cite{Kuroda2017MWSM,Ikhlas2017NernstAFM,Nakatsuji2015LAHEAFM,Tsai2020TAFM}, Co$_3$Sn$_2$S$_2$\cite{liu2019magnetic,morali2019fermi} and Co$_2$MnGa\cite{belopolski2019discovery,Sakai2018NernstFSM}, in which chiral zero Landau levels are expected to exist at magnetic vortex cores or magnetic domain walls. The idea of chiral zero Landau levels induced by pseudo-gauge fields has also been generalized to the strain effect in Dirac/Weyl semimetals~\cite{Chernodub2014AME,Cortijo2015StrainWSM, Grushin2016WSMDSMAxial,Franz2016ChiAnoStrainDSMWSM,Cortijo2016WSMStrain,Chernodub2017DSMChiralAnomaly}.
For example, in \refcite{Franz2016ChiAnoStrainDSMWSM}, it was proposed that torsion in a Dirac/Weyl semimetal nanowire can induce a pseudo-magnetic field and lead to the separation of the left- and right-moving chiral modes to the bulk and surface of the wire. 
Strain-induced pseudo-magnetic field was first experimentally demonstrated in 2D graphene systems~\cite{Levy2010StrainPMF, Guinea2010StrainQHEGraphene}. 
Some experimental evidence of strain induced pseudo-magnetic field has also been found in certain 3D Weyl semimetal systems~\cite{Kamboj2019StrainPMFWSM}.
In addition, significant experimental progress has been achieved in observing chiral zero Landau levels induced by pseudo-gauge field in photonic and acoustic Weyl metamaterials~\cite{Jia2019ChiralZeroModeWeyl, Peri2019StrainWSM}. 
In \refcite{Jia2019ChiralZeroModeWeyl}, it was shown that by engineering the individual unit cells, the pseudo-magnetic field can be created in an inhomogeneous Weyl metamaterial and induce the chiral zero Landau levels. The chiral zero Landau levels are demonstrated experimentally through the observation of one-way propagation of light in this system. More discussion on the experimental progress of the probe of pseudo-gauge field can be found in \refcite{Ilan2020NatRevPseudoGauge}.

\subsection{Magnetic-Resonance-Induced Current Response in Axion Insulators}
\label{sec:MR_AI_ch-2}
In the above part, we have discussed pseudo-magnetic fields induced by the spatial variation of magnetization.
One natural question is whether the temporal variation of magnetization can induce pseudo-electric fields, and if so, what physical phenomena can be induced by pseudo-electric fields. 
In this section, we will switch to the axion insulator, and study how magnetic dynamics influences current response in this phase. 

\begin{figure}[htb]
\centering
\includegraphics[width=\columnwidth]{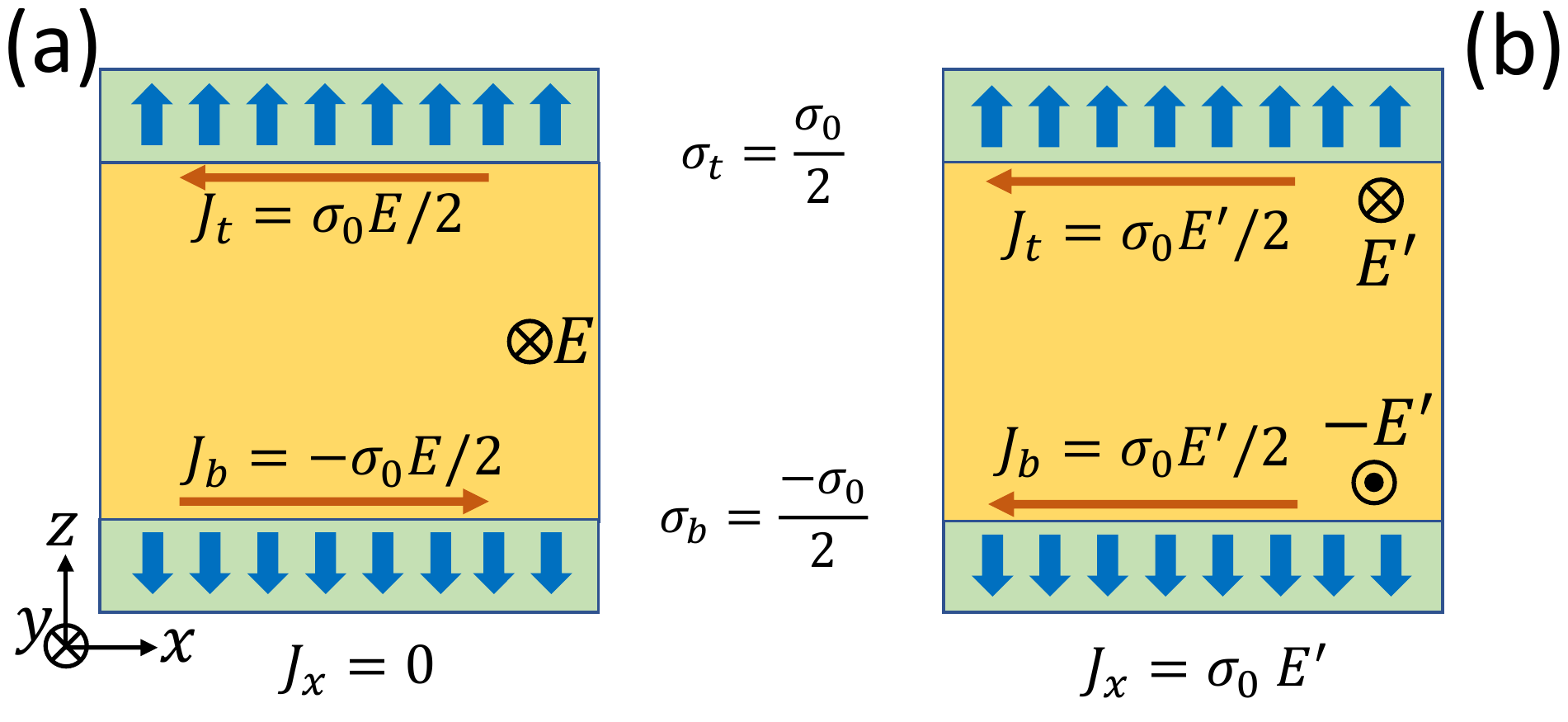}
\caption{\label{fig:setup_AI}
(Adapted from \refcite{Yu2019MRAI}.) In the two graphs, the yellow middle parts are TIs, and the green parts are FMIs with blue arrows the magnetization.
The surface Hall conductances are opposite $\sigma_t=-\sigma_b=\sigma_0/2$ with $\sigma_0=-e^2/(2\pi)$.
In the uniform electric field $E$ along $y$, the axion insulator in (a) has zero total Hall current ($J_x=0$) due to the opposite surface Hall currents ($J_t=-J_b$).
If a pseudo-electric field $E'$ is in opposite $y$ directions on the two surfaces, the axion insulator in (b) has non-zero Hall current.
}
\end{figure}

As mentioned above, a TR-invariant TI host $2+1$D massless Dirac quasi-particles on the surface~\cite{Hasan2010TI,Qi2010TITSC,Qi2008TFT}.
Now let us include a surface magnetic coating with a hedgehog magnetization configuration, which assigns nonzero masses to all the massless Dirac quasi-particles.
In practice, such surface coating is hard to implement, and the common and equivalent setup is the ferromagnetic insulator-TI-ferromagnetic insulator (FMI-TI-FMI) heterostructure~\cite{Wang2015AI,Mogi2017AI,Xiao2018AI} shown in \figref{fig:setup_AI}.
As a result, the effective action of the resultant system for electromagnetic response (response to $\U{1}$ gauge field) contains the following term
\eq{
\label{eq:EdotB_ch-1}
S_{\theta}=-\frac{e^2}{2\pi} \int dt d^3 r \frac{\theta}{2\pi} \bsl{E}\cdot \bsl{B}\ ,
}
which is equivalent to \eqnref{eq:S_eff_a_gen} when the pseudo-gauge field is omitted.
Owing to the nonvanishing effective action field $\theta$, the setup is called the axion insulator.

Although \eqnref{eq:EdotB_ch-1} seems to be a 3+1D action, this action is actually a total derivative in the bulk as $\theta=\pi$ $(\text{mod}\ 2\pi)$ in the bulk, and thus the electromagnetic response of the system is governed by the low-energy surface massive Dirac quasi-particles, since the nonuniform spatial dependence of the static $\theta$ only occurs near the surface. 
Then, we can focus on the response of surface Dirac quasi-particles in FMI-TI-FMI heterostructure under an uniform external electric field.
As schematically shown in \figref{fig:setup_AI}(a), the applied electric field cannot induce any current response, since the top and bottom massive Dirac quasi-particles have opposite Hall conductance and thus have opposite Hall currents that cancel each other. 
This means that there will be a zero-Hall plateau, which has been found in magneto-transport experiments~\cite{Mogi2017AI,Xiao2018AI}. 
However, measuring the zero-Hall plateau is not a conclusive evidence for the axion insulator phase since the normal insulator also has zero Hall resistance.
The conclusive signature for the axion insulator is the topological magnetoelectric effect shown in \eqnref{eq:TME_bulk} (after omitting the pseudo-electric field).
In \figref{fig:setup_AI}(a), the effect corresponds to the magnetization directly given by the counter-propagating Hall currents of the massive Dirac quasi-particles, which is still challenging to measure in experiments. 

For observing a nontrivial signature of the axion insulator, we can try to induce a nonzero current response by applying a pseudo-electric field that points to opposite directions for the two massive Dirac quasi-particles \figref{fig:setup_AI}(b).
As a result, the two surfaces would have the same Hall currents and add to a nonzero value.
\refcite{Yu2019MRAI} suggests that the desired pseudo-electric field can be induced by the dynamical magnetization of the FMIs, as discussed below.

The Hamiltonian for the two massive Dirac quasi-particles reads~\cite{Garate2010ISGE}
\eq{
H=\int \frac{d^2 k}{(2\pi)^2}\sum_{i} c^\dagger_{\bsl{k},i} [h_{0,i}+h_{Z}+h_{ex,i}] c_{\bsl{k},i}\ .
}
\eq{
h_{0,i}=v_{f,i}[- \sigma_y ( k_x+e A_{x})+ \sigma_x (k_y+e A_{y})]+(-e) \varphi
}
depicts the surface Dirac quasi-particles $c^\dagger_{\bsl{k},i}=(c^\dagger_{\bsl{k},i,\uparrow},c^\dagger_{\bsl{k},i,\downarrow})$ coupled to a 2+1D $\U{1}$ gauge field $A^\mu=(\varphi,A_{x},A_{y})$, where $i=t,b$ labels the top and bottom surfaces, respectively, $\bsl{k}=(k_x,k_y)$, $v_{f,t}=-v_{f,b}=v_f$ and $\sigma_{x,y,z}$ are Pauli matrices for spin.
$h_{ex}=g_M \bsl{M}_i\cdot \bsl{\si}$ is the exchange coupling term with $g_M$ assumed to be positive and the same on both surfaces for simplicity.
Since we care about the magnetic resonance, we would add uniform magnetic field $\bsl{B}$, which induces Zeeman term $h_{Z}=\mu_B \bsl{B}\cdot \bsl{\si}$ with $\mu_B$ Bohr magneton.

In order to compare with the action formalism used in the previous sections, we rewrite the Hamiltonian into the action form
\eq{
\label{eq:S_ch-2_sec-3}
S=\int d^3 x \sum_i \bar{\psi}_{i}[\ii \Gamma^\mu_i (\partial_\mu+ \ii e A_{\mu}+ \ii e A^{pse}_{i,\mu})- m_{i,z} ]\psi_{i}
}
where $x^\mu=(t,x,y)$, $\Gamma^\mu_i=(\sigma_z,v_{f,i}\ii \sigma_x,v_{f,i} \ii \sigma_y)$, and
\eq{
\label{eq:A_pse_ch-2_sec-3}
A^{pse}_{i,\mu}=\frac{1}{e v_{f,i}}(0, - m_{i,y}, m_{i,x} )_{\mu}
}
with $\bsl{m}_{i}=\mu_B \bsl{B}+g_M \bsl{M}_{i}$.
Compared to \eqnref{eq:Dirac_QP_A_Apse}, we can see both the external uniform magnetic field and the magnetization in FMIs can contribute to the pseudo-gauge field.
Another difference between \eqnref{eq:S_ch-2_sec-3} and \eqnref{eq:Dirac_QP_A_Apse} is that we allow arbitrary velocity in  \eqnref{eq:S_ch-2_sec-3}, which in principle can be canceled by a proper scaling.

In particular, even if the magnetization and magnetic field have the same $x,y$ components for two Dirac quasi-particles, the induced pseudo-electric field would point to the opposite directions for the two Dirac quasi-particles at the opposite surfaces owing to $v_{f,t}=-v_{f,b}$.
As suggested by \figref{fig:setup_AI}(b), such a pseudo-gauge field should induce a nonzero current response.
To derive the current response induced by the pseudo-gauge field, let us consider a uniform magnetic field that only has an oscillating $x$ component, \ie, $\bsl{B}(t)=(B_0 \cos(\omega t),0,0)$ with the constant $B_0$.
By integrating out the fermionic modes in \eqnref{eq:S_ch-2_sec-3}, the current response along $x$ direction to the leading order can be obtained as
\eq{
\label{eq:J_x_AP_ch-2_sec-3}
J^x_{AP} =J_B + J_M\ ,
}
where
\eq{
J_B=\sigma_t B_0 \omega \sin(\omega t) (- L_z+\frac{2}{e v_f} \mu_B )
}
is the current density induced by the external magnetic field, $L_z$ is the distance between two surfaces, and 
\eq{
J_M= -\frac{\si_t}{e v_f} g_M(\dot{M}_{t,x}+\dot{M}_{b,x})
}
is given by the magnetization.
When tuning the frequency to induce the ferromagnetic resonance, \refcite{Yu2019MRAI} shows that the magnetization-induced $J_M$ can be much larger than $J_B$.
Specifically, with typical values of parameters, the ratio between the maximum values of $J_M$ and $J_B$ is about $20\sim 2\times 10^4$ depending on the material for FMIs.
For the same parameter values, the current amplitude induced by the magnetization is typically $10\sim 10^4$nA, which enters the experimentally testable range.

The axion insulator phase has also been predicted in anti-ferromagnetic TI \mbt~\cite{Zhang2018AIMnBi2Te4} and various other materials as comprehensively reviewed in \refcite{Sekine2020AEMaterials}.
In \mbt, \refcite{Yu2019MRAI} suggests the pseudo-gauge field can be given by  anti-ferromagnetic resonance and further induces nonzero current response, through a similar mechanism.
Besides the pseudo-gauge field mechanism, dynamical magnetization can also make the axion field dynamical and then induce a nonzero response of axion insulators~\cite{Wang2020AnisotropicTME,Liu2020DynamicalAxionMR}.
The axial magnetoelectric effect was also studied in the presence of the dynamical pseudo-gauge fields~\cite{Liang2020AxialMagnetoelectricDSM}.

\section{Strain-Induced Pseudo-gauge Field and Piezo-electromagnetic Response}

\label{sec:strain_pseudogauge}

\begin{figure}[t]
    \centering
    \includegraphics[width=\columnwidth]{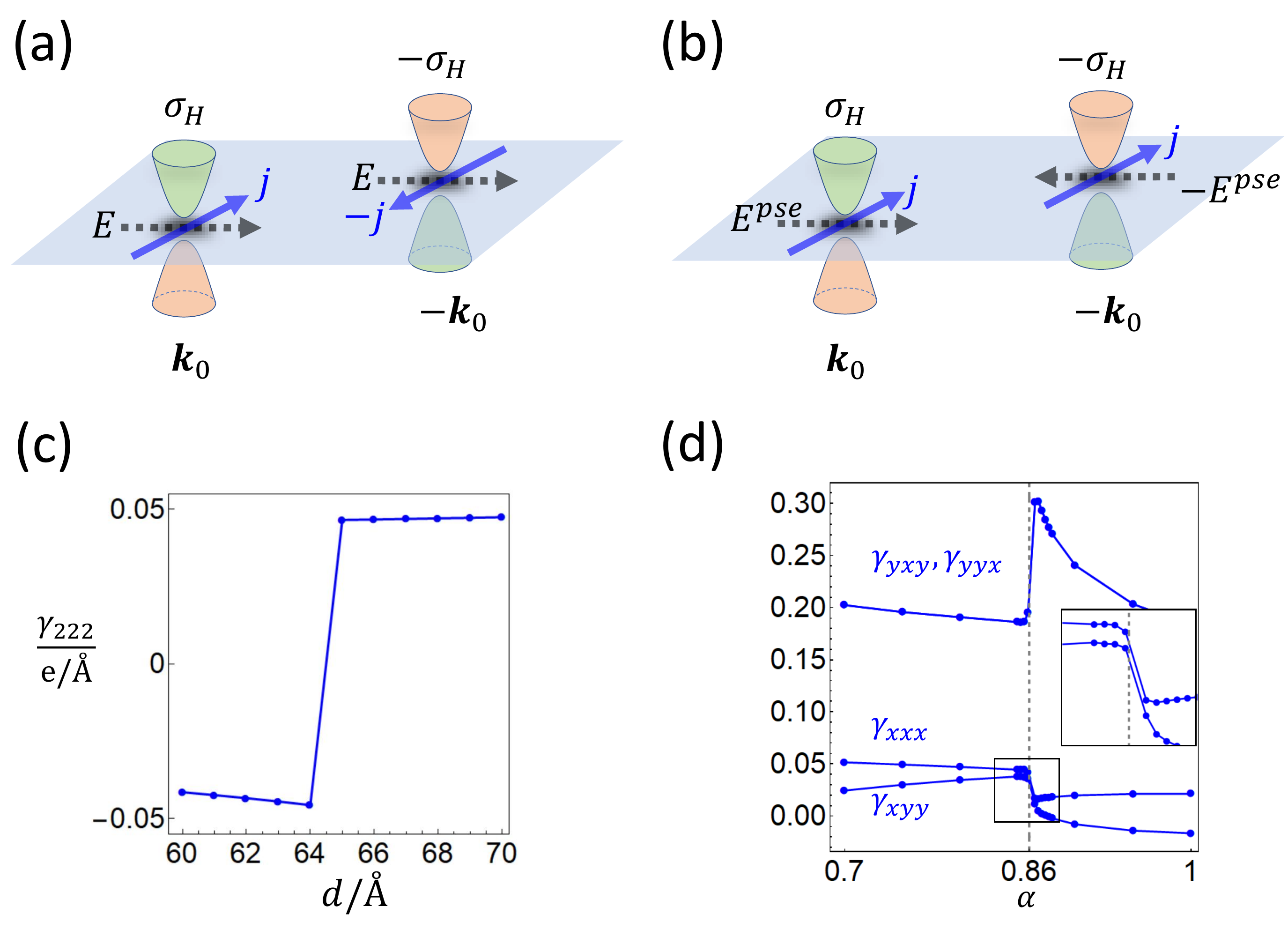}
    \caption{
    (Adapted from \refcite{Yu2019PETTQPT} and \refcite{Yu2020DPME}.)
    2+1D response effects in TR-invariant insulators with two massive Dirac quasi-particles.
    In (a) and (b), two 2+1D massive Dirac quasi-particles at $\pm \bsl{k}_{0}$ are related by TR symmetry.
    The two massive Dirac quasi-particles provide opposite and canceling contributions to the Hall conductance $\sigma_H$.
    In (c), the piezoelectric tensor component $\gamma_{222}$ of the HgTe quantum well is plotted as the function of the thickness $d$.
    The $\dsZ_2$ topological quantum phase transition happens around $d=65\AA$.
    In (d), the piezoelectric tensor components of BaMnSb$_2$ are plotted as the function of the distortion strength $\alpha$.
    The $\dsZ_2$ topological quantum phase transition happens around $\alpha=0.86$.
    }
    \label{fig:PET}
\end{figure}

The previous section focused on the pseudo-gauge fields induced by magnetization and the corresponding physical phenomena. An alternative way to create pseudo-gauge fields is through static or dynamic strain. A number of early theoretical studies~\cite{Chernodub2014AME,Cortijo2015StrainWSM,Grushin2016WSMDSMAxial,Franz2016ChiAnoStrainDSMWSM,Cortijo2016WSMStrain,Chernodub2017DSMChiralAnomaly,Antebi2020AnomalyWSM} explored the chiral anomaly due to the strain-induced pseudo-gauge fields in Dirac/Weyl semimetals, as mentioned above in \secref{sec:mag_pseudogauge}, and various other aspects of the strain-induced pseudo-gauge field were also discussed in a number of literature ~\cite{Zhou2013StrainInducedPGF,Sumiyoshi2016TorsionalWSM,Gorbar2017WSMChiralPlasmons,Gorbar2017ChiralPlasmons,Gorbar2017Pseudomagnetic,Arjona2018StrainWSM,Matsushita2018CryDefWDSM,Kobayashi2018WSC,Ishihara2019SkyrmionWSF,Heidari2019HallPhonon,Chernodub2019ChiralSoundWSM,Behrends2019WSMStrain,Behrends2019LatticeCA,Ferreiros2019AxialTosionalWSM,Matteo2020AxialGF,Ferreiros2020ChiralResTorsion,Hannukainen2020AxialWSM,Hannukainen2020ElectricAxialWSM,Matsushita2020PiezoNLSM,Ghosh2020PlanarHallWSM}.
In this section, we mainly review our works on the piezo-electromagnetic response induced by pseudo-electric field, including the jump of piezo-electric tensor across a 2D TR-invariant topological phase transition and the dynamical piezo-magnetic effect in 3D Weyl semimetals gapped by a CDW.

\subsection{2+1D Piezoelectric Effect}

The strain-induced pseudo-gauge field for 2+1D Dirac quasi-particles was first studied in graphene~\cite{Guinea2010PGFGraphene,Guinea2010StrainQHEGraphene,Levy2010StrainPMF} with a focus on the resultant pseudo-magnetic field. 
Later on, the strain-induced pseudo-electric field and the resultant piezoelectric effect was discussed in graphene with staggered potential~\cite{Vaezi2013StrainGraphene}.
The relation between pseudo-gauge field and the piezoelectric effect~\cite{Martin1972Piezo,Vanderbilt2000BPPZ} was also discussed in systems like h-BN~\cite{Droth2016PETBN,Rostami2018PiE} and monolayer transition metal dichalcogenides~\cite{Rostami2018PiE}. (See a detailed review in \refcite{Amorim2016StrainPGFGraphene}.)

In this section, we will first review the intuitive picture for the relationship between pseudo-gauge field and piezoelectric effect.
We start with a 2+1D TR-invariant insulator whose low-energy physics is captured by two 2+1D massive Dirac quasi-particles (also known as two valleys), as shown in \figref{fig:PET}(a-b).
The story is similar to \figref{fig:setup_AI} but the two Dirac quasi-particles now live in a bulk of the 2+1D insulator, instead of surfaces of 3+1D system.
The TR symmetry relates one massive Dirac quasi-particle to the other one, and thereby the two Dirac quasi-particles would have opposite Hall conductance (\ie, opposite valley Hall conductance).
As a result, applying a uniform electric field induces Hall currents for the two Dirac quasi-particle, which point in opposite directions and cancel (\figref{fig:PET}(a)).
On the other hand, applying a dynamical homogeneous strain can induce pseudo-electric fields that points to opposite directions for the two Dirac quasi-particles, and the piezoelectric field $E^{pse}$ is roughly proportional to the time derivative of the strain tensor $u$, \ie, $E^{pse}\sim\dot{u}$.
As the opposite directions of pseudo-electric fields cancel the opposite signs of the valley Hall conductance (\figref{fig:PET}(b)), the pseudo-electric fields drive a nonzero total Hall current density $J\sim \dot{u}$.
The nonzero current response driven by time directive of homogeneous strain is the piezoelectric effect, owing to the following expression of the piezoelectric tensor $\gamma_{ijk}$~\cite{Martin1972Piezo,Vanderbilt2000BPPZ}
\eq{
\label{eq:PET_Pro}
\gamma_{ijk}=
\left.\frac{\partial J_i}{\partial \dot{u}_{jk}}\right|_{u_{jk},\dot{u}_{jk}\rightarrow 0}\ ,
}
where $i,j,k=x,y$, and $\bsl{u}$ is the displacement at $\bsl{x}$.
$u_{ij}=\partial_{x_i} u_j$ is the displacement gradient, and its symmetric part $(u_{ij}+u_{ji})/2$ is the strain tensor, whereas the anti-symmetric part $(u_{ij}-u_{ji})/2$ is the rotation.
If only strain is considered (like \refcite{Yu2019PETTQPT}), then only the symmetry part of $u_{ij}$ enters the Hamiltonian, resulting in $\gamma_{ijk}=\gamma_{ikj}$.

Despite the contribution from the low-energy Dirac quasi-particles, high-energy modes can also contribute to the piezoelectric effect of an insulating TR-invariant Dirac materials.
Therefore, it is hard to experimentally distinguish from the trivial background the Dirac part proposed in the early theoretical works.
To resolve this issue, \refcite{Yu2019PETTQPT} proposed to focus on the behavior of the piezoelectric effect across a topological quantum phase transition instead of that in a single insulating phase.
To be more specific, it was theoretically predicted in \refcite{Yu2019PETTQPT} that the piezoelectric effect has a discontinuous change across a topological quantum phase transition, and the discontinuous change solely comes from the Dirac quasi-particles without involving the high-energy modes.
In the following, we will discuss a simple case to illustrate the main idea.
Interested readers may resort to \refcite{Yu2019PETTQPT} for more details.

Let us consider a 2+1D TR-invariant spin-orbit coupled system with no crystalline symmetries other than the lattice translation (plane group $p1$).
Low-energy physics of such system is typically captured by two TR-related Dirac quasi-particles (\figref{fig:PET}(a-b)).
Instead of using the action form in \secref{sec:MR_AI_ch-2}, we use an equivalent Hamiltonian-based formalism here, in order to show that the Dirac physics can be captured in both ways.
The two Dirac quasi-particles can be described by the following effective matrix Hamiltonian
\eqa{
\label{eq:h_pm_0_p1_sim}
 h_{s,0}(\bsl{q})= v_x q_x\sigma_x+v_y q_y\sigma_y + s m \sigma_z\ ,
}
where $\bsl{q}$ labels the momentum of the Dirac quasi-particles, $s=\pm$ labels the two Dirac quasi-particles, the TR symmetry is represented as $\TR\dot{=} \ii\sigma_y \cc$ with $\cc$ the complex conjugate, and $\sigma$'s are Pauli matrices.
\eqnref{eq:h_pm_0_p1_sim} is a simplified version of the most general form, and we choose \eqnref{eq:h_pm_0_p1_sim} just to illustrate the main physics.
The mass $m$ of the Dirac quasi-particle with $s=+$ is the key tuning parameter for the topological quantum phase transition.
When tuning $m$ from positive to negative, the system undergoes a $\dsZ_2$ topological quantum phase transition that separates the $\dsZ_2$ TI (also called quantum spin Hall insulator) phase and the normal insulator phase.

Now let us introduce the electron-strain coupling around based on the TR symmetry:
\eq{
\label{eq:h_+-_1_p1}
h_{\pm,1}(u)=\pm \sum_{i,j}(\xi_{x,ij}\sigma_{x}+\xi_{y,ij}\sigma_{y}) u_{ij}\ ,
}
where $\xi$'s are the electron-strain coupling that is related to the electron-phonon coupling~\cite{Suzuura2002Phonon-Electron}.
We have neglected the electron-strain couplings that involves $\sigma_0$ and $\sigma_z$, since they preserve an effective inversion symmetry of \eqnref{eq:h_pm_0_p1_sim} and thereby cannot contribute to the piezoelectric effect.
Combining \eqnref{eq:h_+-_1_p1} with \eqnref{eq:h_pm_0_p1_sim}, we arrive at the total Hamiltonian $h_\pm(\bsl{q},u)=h_{\pm,0}(\bsl{q})+h_{\pm,1}(u)$ as
\eqa{
\label{eq:A_pse_p1}
h_\pm(\bsl{q},u) =\left[v_x (q_x\pm A_x^{pse})\right]\sigma_x+\left[v_y (q_y\pm A_y^{pse})\right]\sigma_y \pm m \sigma_z\ ,
}
where $A_x^{pse}=\sum_{i,j}\xi_{x,ij}u_{ij}/v_x$ and $A_y^{pse}= \sum_{i,j}\xi_{y,ij}u_{ij}/v_y$.

The piezoelectric tensor for the Hamiltonian (\ref{eq:A_pse_p1}) can be directly calculated from the following expression~\cite{Vanderbilt2000BPPZ,Wang2018TBPiezo}
\eqa{
\label{eq:gamma_bc}
\gamma_{ijk}^{eff}=&-e\int \frac{d^2 q}{(2\pi)^2} \sum_{s}\left. F_{q_i, u_{jk}}^{s}\right|_{u_{jk}\rightarrow 0}\ ,
}
where $F_{q_i, u_{jk}}^{\pm}$ term has a Berry-curvature-like expression as
\eq{
\label{eq:F}
F_{q_i, u_{jk}}^{\pm}=(-\ii)\left[\bra{ \partial_{q_i} \varphi_{\pm,\bsl{q}}} \partial_{u_{jk}} \varphi_{\pm,\bsl{q}}\rangle-( \partial_{q_i}\leftrightarrow \partial_{u_{jk}})\right]
}
with $\ket{\varphi_{\pm,\bsl{q}}}$ the periodic part of the occupied Bloch state for the Dirac quasi-particles in the presence of the strain, and $eff$ means that the tensor solely comes from the Dirac quasi-particles.
As suggested by \eqnref{eq:A_pse_p1}, $u_{jk}$ always appears in the matrix Hamiltonian as $q_i\pm A_i^{pse}$, and thereby it would appear in the same form in the representation of $\ket{ \varphi_{\pm,\bsl{q}} }$, resulting in 
\eq{
\partial_{u_{ij}} \ket{\varphi_{\pm,\bsl{q}}}=\frac{\partial A_{i'}^{pse}}{\partial u_{ij}}\partial_{A_{i'}^{pse}} \ket{\varphi_{\pm,\bsl{q}}}=\pm \frac{\partial A_{i'}^{pse}}{\partial u_{ij}}\partial_{q_{i'}} \ket{\varphi_{\pm,\bsl{q}}}\ .
}
The above expression would transform \eqnref{eq:F} into the conventional Berry curvature, and we arrive at the following expression for the piezoelectric tensor
\eqa{
\label{eq:PET_F}
&\gamma_{xij}^{eff}=-\frac{e}{2\pi} \sum_{s=\pm} s N_s \frac{\partial A_{y}^{pse}}{\partial u_{ij}} \\
&\gamma_{yij}^{eff}=\frac{e}{2\pi} \sum_{s=\pm} s N_s \frac{\partial A_{x}^{pse}}{\partial u_{ij}}\ .
}
$N_s$ in \eqnref{eq:PET_F} is the valley Chern number of the Dirac quasi-particle with the index $s$, which is the integral of the conventional Berry curvature $F^s_{xy}(\bsl{q})$ at zero strain
\eq{
N_s=\int \frac{d^2q}{2\pi} F^s_{xy}(\bsl{q})\ .
}
When tuning the mass $m$ from negative to positive, a topological quantum phase transition will occur and lead to the change in piezoelectric tensor, that would be proportional to the change of the valley Chern number as
\eqa{
\label{eq:PET_jump_p1}
&\Delta\gamma_{xij}=-e\frac{\Delta N_+}{\pi} \frac{\xi_{y,ij}}{v_y} \\
&\Delta\gamma_{yij}=e\frac{\Delta N_+}{\pi} \frac{\xi_{x,ij}}{v_x} \ ,
}
where $\Delta N_+ = -\sgn{v_x v_y}$.
Since the $v_x$, $v_y$, and $\xi$'s are typically nonzero, the jump of the piezoelectric tensor given by Dirac quasi-particles is nonzero.
On the other hand, all high-energy modes undergo an adiabatic evolution across the transition, and thereby their contribution to the piezoelectric tensor should not have any discontinuous change.

As discussed in \refcite{Yu2019PETTQPT}, such discontinuous changes of piezoelectric tensor happen for all codimension-1 transitions and for all 7 plane groups that allow the piezoelectric effect, in the presence of TR symmetry and spin-orbit coupling.
Furthermore, the piezoelectric jump was predicted to appear in HgTe quantum well upon tuning the thickness (\figref{fig:PET}(c)), and in the quasi-two-dimensional BaMnSb$_2$ upon tuning the distortion (\figref{fig:PET}(d)).

\subsection{Dynamical Piezomagnetic Effect}

The above subsection shows the strain-induced pseudo-gauge field for 2+1D Dirac quasi-particle and its dramatic influence on the piezoelectric effect. 
A natural question immediately follows: can strain induce pseudo-gauge field in 3D gapped Dirac materials, and if so, what is the resultant phenomenon?
\refcite{Yu2020DPME} answered this question by showing that strain indeed can act as a pseudo-gauge field for 3+1D Dirac quasi-particles and the resultant effect is the dynamical piezomagnetic effect (DPME), which will be reviewd in this part.
In this part, we will only show an intuitive and simplified discussion, while a comprehensive and more rigorous discussion can be found in \refcite{Yu2020DPME}.

\begin{figure}[t]
    \centering
    \includegraphics[width=\columnwidth]{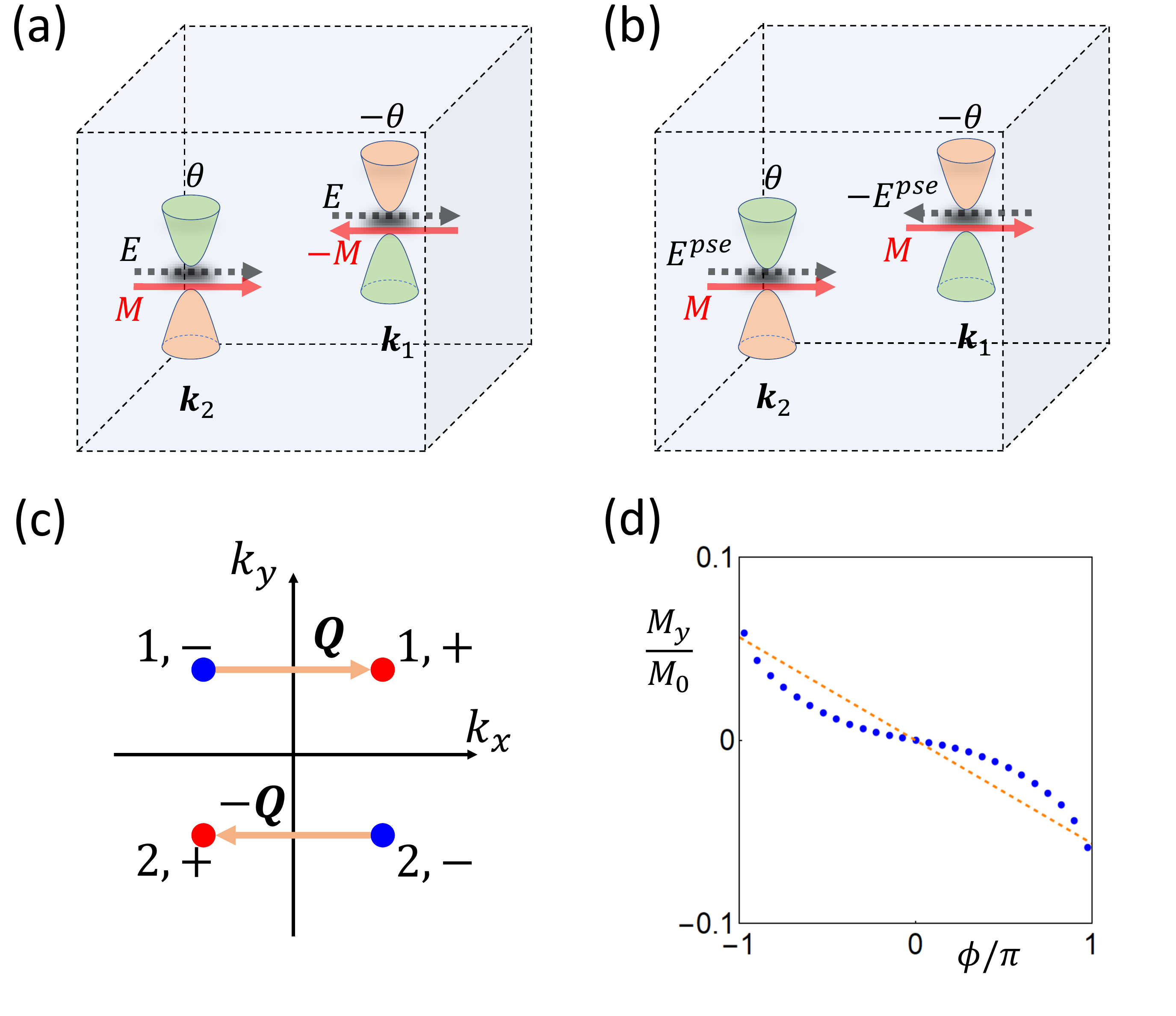}
    \caption{
    (Adapted from  \refcite{Yu2020DPME}.)
    3+1D DPME in TR-invariant insulators with two massive Dirac quasi-particles.
    In (a) and (b), two 3+1D massive Dirac quasi-particles at $\bsl{k}_{1,2}$ are related by TR symmetry.
    Owing to TR symmetry, the two massive Dirac quasi-particles have complex masses and have opposite axion fields $\theta$.
    (c) Four Weyl points in a TR-invariant minimal model of Weyl semimetal.
    The four Weyl points are projected onto the $k_x-k_y$ plane, and the arrows indicate the projection of the CDW wavevectors $\bsl{Q}$.
    (d) The plot of bulk magnetization along $y$ as a function of the phase angle $\phi$ of the CDW order parameter.
    The blue dots and orange dashed line indicate data obtained from the tight-binding calculation and the effective action, respectively.
    $M_0=e \dot{u}_{zz} /a_0$ with $a_0$ the lattice constant of the tight-binding model before introducing CDW.
    }
    \label{fig:DPME}
\end{figure}

Let us consider a TR-invariant 3D insulator with two 3+1D Dirac quasi-particles, as shown \figref{fig:DPME}(a-b).
Here we consider the Dirac quasi-particle with the complex masses, and its Lagrangian is described by \eqnref{eq:Dirac_QP_A_Apse} with $a=1,2$, while neglecting the chiral gauge field $A_5$ and imposing $m_a=|m_a|$.
Owing to the TR symmetry, $\theta_a$ satisfies $\theta_2=-\theta_1\equiv \theta$.
When we apply a uniform electric field $\bsl{E}$, the vector potential $\bsl{A}$ is spatially uniform but time-dependent owing to $\bsl{E}=d\bsl{A}/dt$. 
Furthermore, we assume the pseudo-gauge field is also spatially uniform, which provides a uniform pseudo-electric field as $\bsl{E}^{pse}_a=d\bsl{A}^{pse}_a/dt$.
Then, from the effective action \eqnref{eq:S_eff_a_gen}, we can derive the bulk magnetization \eqnref{eq:TME_bulk}.
Clearly, owing to the opposite $\theta_a$ for the two Dirac quasi-particles, the magnetizations induced by a uniform electric field would be opposite for the two Dirac quasi-particles and cancel (\figref{fig:DPME}(a)), while the magnetizations induced by pseudo-electric fields that satisfy $\bsl{E}^{pse}_{1}=-\bsl{E}^{pse}_{2}$ would point in the same direction for two Dirac quasi-particles and add to a nonzero total value (\figref{fig:DPME}(b)).

As proposed in \refcite{Zhang2013WSMCDWAxion,Roy2015WSMCDWAxion,Gooth2019WSMCDWAxion,Yu2020DPME}, the above-discussed TR-related 3+1D Dirac quasi-particles with complex masses exist in a TR-invariant Weyl semimetal gapped by a CDW.
To see this, let us focus on a minimal TR-invariant Weyl semimetal with four Weyl points (\figref{fig:DPME}(c)) at
\eq{
\label{eq:WP_pos}
\bsl{k}_{a,\alpha}=(-1)^{a-1}(\alpha k_{0,x}, k_{0,y}, \alpha k_{0,z})\ ,
}
where $\alpha=\pm$ indicates the relative chirality of the Weyl points.
Under TR-symmetry, $\bsl{k}_{1,\alpha}$ and $\bsl{k}_{2,\alpha}$ are interchanged; we also impose a mirror symmetry $m_y$ (perpendicular to $y$ direction) which relates $\bsl{k}_{1,\alpha}$ to $\bsl{k}_{2,-\alpha}$.
We introduce mirror symmetry just for simplicity, and the underlying physics do not rely on it.
With this convention, the low-energy Lagrangian for the Weyl points are
\eq{
\mathcal{L}_{WP}=\sum_{a,\alpha}\psi^\dagger_{a,\alpha}\left[ \ii \partial_t-\alpha\sum_{i=x,y,z} (-\ii\partial_{i}-k_{a,\alpha,i}) \sigma_i\right]\psi_{a,\alpha}
\ ,
}
where $\psi^\dagger_{a,\alpha}$ is a two-component fermionic field for the Weyl point with the indices $(a,\alpha)$, and the representations of TR and mirror symmetries for the Weyl field are $\ii\sigma_y \cc$ and $-\ii\sigma_y$, respectively.
In the following, we will neglect $k_{a,\alpha,i}$ since it is not related to the DPME.
By defining $\gamma^\mu=(\tau_x\sigma_0,-\ii \tau_y \bsl{\sigma})_\mu$ and $\psi_{a}=(\psi_{a,+},\psi_{a,-})^T$, the total action of free Weyl quasi-particles can be rewritten as \eq{
\label{eq:L_WP}
\mathcal{L}_{WP}= \sum_{a} \overline{\psi}_a \ii \slashed{\partial}\psi_{a}\ ,
}
where $\overline{\psi}_a=\psi^\dagger_{a}\gamma^0$. 

Next, we include a bulk-uniform symmetry-preserving mean-field CDW order parameter whose wave-vector $\bsl{Q}=\bsl{k}_{1,+}-\bsl{k}_{1,-}=-(\bsl{k}_{2,+}-\bsl{k}_{2,-})$ is shown in \figref{fig:DPME}(c).
As shown in \figref{fig:DPME}(c), the order parameter couples two Weyl points with the same $a$.
After including the CDW, the action of Weyl points becomes~\cite{Zhang2013WSMCDWAxion}
\eq{
\label{eq:L_WP_CDW_MF}
\mathcal{L}_{WP+CDW}=\sum_{a} \overline{\psi}_a (\ii \slashed{\partial}-|m| e^{-\ii (-1)^{a-1} \phi \gamma^5} )\psi_{a}\ ,
}
where $|m|$ is CDW magnitude, $\phi$ is the phase angle of the CDW order parameter, and we have neglect the CDW wave-vector related term for simplicity.
Comparing \eqnref{eq:L_WP_CDW_MF} to \eqnref{eq:Dirac_QP_A_Apse}, we can indeed see two Weyl points coupled by the CDW order parameter form one Dirac quasi-particle, and the CDW order parameter provides the complex mass for the Dirac quasi-particle.
In particular, 
\eq{
\theta_a=(-1)^{a-1} \phi
}
is the effective valley axion field.

We further include the following symmetry-allowed electron-strain couplings of interest
\eqa{
\label{eq:L_str}
&\mathcal{L}_{str}=\sum_{a}\overline{\psi}_{a} [-\xi_0\gamma^0+(-1)^{a}\gamma^2 \xi_y]\psi_{a}u_{zz} \ .
}
Combining \eqnref{eq:L_str} and \eqnref{eq:L_WP_CDW_MF} and including the $\U{1}$ gauge field, we will arrive at the \eqnref{eq:Dirac_QP_A_Apse} with $A_{a,5}=0$, $m_a=|m|$ and
\eq{
A^{pse}_{a,\mu}=\frac{u_{zz}}{e}(\xi_0 , 0, (-1)^{a-1}\xi_y ,0)_\mu\ ,
}
The $y$ components of the pseudo-gauge fields $A^{pse}_{a,y}$ have opposite signs for the two Dirac quasi-particles, giving rise to the following pseudo-electric field
\eq{
\label{eq:E_pse}
\bsl{E}^{pse}_a=(-1)^a \frac{\xi_y}{e} \dot{u}_{zz} \bsl{e}_y\ .
}
As a result, the total magnetization reads
\eq{
\label{eq:M_bulk}
\bsl{M}=-\sum_a \frac{e^2}{2 \pi}\frac{\theta_a}{2\pi} \bsl{E}^{pse}_{a}=\frac{e \xi_y }{2\pi^2 }\phi \dot{u}_{zz} \bsl{e}_y\ .
}
According to \eqnref{eq:M_bulk}, the magnetization is proportional to the time-derivative of the strain tensor, making the effect different from the conventional piezomagnetic effect where the magnetization is proportional to the strain tensor itself.
Therefore, we call it DPME.

As shown by the dashed line in \figref{fig:DPME}(d), the magnetization would have a discontinuous change upon varying $\phi$ from $-\pi+0^+$ to $\pi+0^-$.
However, at the meantime, the bulk action \eqnref{eq:L_WP_CDW_MF} undergoes a continuous change. This implies that the discontinuous change does not origin from the bulk of the system.
Indeed, \refcite{Yu2020DPME} performed a tight-binding calculation that reproduces the discontinuous change of DPME (blue dots in \figref{fig:DPME}(d)), and verified that the change is induced by surface topological quantum phase transition~\cite{Khalaf2019BoundaryObstructedTopo}.
Furthermore, the response coefficient for the DPME with reasonable parameter values is estimated to be  $|\partial\bsl{M}/\partial\dot{u}_{zz}|\sim 0.8 e/\AA$.
The coefficient turns out to have the same units and the same order of magnitude as the experimentally observable 2+1D piezoelectric coefficient~\cite{Zhu2014PET}. Thus, the DPME should be experimentally testable.

At last, we emphasize that all previous studies on topology-related strain-induced phenomena in gapped systems focused on 2D Berry curvature, while the response coefficient of DPME in \refcite{Yu2020DPME} is determined by the effective axion field, whose bulk average value is given by Chern-Simons 3-form.
As the Chern-Simons 3-form cannot exist in 2D (or even lower dimensional) manifold, the DPME is an intrinsic 3D phenomenon that can only exist in 3D (or higher spacial dimensions), and thereby is fundamentally different from all previous strain-induced phenomena determined by the 2D Berry curvature.

\section{Conclusion}
\label{sec:conclusion}
In sum, we have described our works in understanding the physical phenomena induced by the magnetization and strain in different types of Dirac/Weyl materials, showing that all these physical phenomena can be understood under the concept of pseudo-gauge fields. 
We emphasize that the concept of pseudo-gauge fields can be applied to various other physical phenomena. For example, phonons can also induce internal strain and thus may play the role of pseudo-gauge fields for Dirac/Weyl quasi-particles through electron-phonon interactions. This viewpoint implies non-trivial phonon dynamics induced by electron-phonon interactions in Dirac/Weyl materials. Moreover, we have only discussed the linear response due to the pseudo-gauge fields, while non-linear response of pseudo-gauge fields is an unexplored direction.
The direction is quite interesting and important, given the recent theoretical development of non-linear topological electromagnetic response in Dirac/Weyl materials~\cite{Morimoto2016NonlinearOptical,DeJuan2017WSMPhotogalvanic,Sukhachov2020PhononPseudogauge}.

On the experimental sides, the pseudo-gauge field has been firmly demonstrated in Dirac/Weyl photonic and acoustic metamaterials~\cite{Jia2019ChiralZeroModeWeyl,Peri2019StrainWSM}. 
For the condensed matter systems, it was suggested that several interface effects, such as Fermi arcs of Dirac/Weyl semimetals, can be understood in the context of pseudo-magnetic field~\cite{Grushin2016WSMDSMAxial}.
However, it is more desirable to realize bulk physical phenomena that are related to the bulk pseudo-gauge field, which is still challenging. A review of the current progress of this topic can be found in \refcite{Ilan2020NatRevPseudoGauge}.
Beside the pseudo-gauge field, the external perturbations can also influence the Fermi velocities of Dirac/Weyl fermions and thus serve as the artificial gravitational field. This may lead to a new field of studying Dirac/Weyl fermions in the artificial gravitational field in Dirac/Weyl materials~\cite{Gooth2019WSMCDWAxion}.

\section{Acknowledgement}
J.~Yu is supported by the Laboratory for Physical Sciences. C.X. Liu is supported by the support of the Office of Naval Research (Grant No. N00014-18-1-2793) and Kaufman New Initiative research Grant No. KA2018-98553 of the Pittsburgh Foundation. 

\bibliography{bibfile_references}

\end{document}